\documentclass[pra,twocolumn,showpacs,superscriptaddress,a4paper]{revtex4}
\usepackage{bm,graphicx,amsmath}

\begin{document}

\title{Polarization phase gate with a tripod atomic system}

\author{S. Rebi\'{c}}
\email[E-mail: ]{stojan.rebic@unicam.it}
\author{D. Vitali}
\author{C. Ottaviani}
\author{P. Tombesi}
\affiliation{INFM and Dipartimento di Fisica, Universit\`{a} di
Camerino, I-62032 Camerino, Italy}
\author{M. Artoni}
\affiliation{INFM and Dipartimento di Chimica e Fisica dei Materiali, Universit\`{a} di Brescia, I-25133
Brescia, Italy}
\affiliation{European Laboratory for Non-linear Spectroscopy, via N. Carrara, I-50019 Sesto
Fiorentino, Italy}
\author{F. Cataliotti}
\affiliation{European Laboratory for Non-linear Spectroscopy, via N. Carrara, I-50019 Sesto Fiorentino, Italy}
\affiliation{INFM and Dipartimento di Fisica, Universit\`{a} di Catania, I-95124 Catania, Italy}
\author{R. Corbal\'{a}n.}
\affiliation{Departament de F\'{i}sica, Universitat Aut\`{o}noma de
Barcelona, E-08193, Bellaterra, Spain}

\date{\today}

\begin{abstract}
We analyze the nonlinear optical response of a four-level atomic system driven into a tripod configuration. The
large cross-Kerr nonlinearities that occur in such a system are shown to produce nonlinear phase shifts of order
$\pi$. Such a substantial shift may be observed in a cold atomic gas in a magneto-optical trap where it could be
feasibly exploited towards the realization of a polarization quantum phase-gate. The experimental feasibility of
such a gate is here examined in detail.
\end{abstract}


\maketitle

\section{Introduction \label{intro}}

A great effort has recently gone into the search for practical architecture for quantum information processing
systems. While most attention has been devoted toward theoretical issues, several strategies have also been
proposed for experimental investigations. However, the laboratory demand for building quantum information
devices are quite severe, requiring strong coupling between \emph{qubits}, the quantum carriers of information,
in an environment with minimal dissipation. For this reason experimental progress has so far lagged behind the
remarkable development that quantum information theory now witnesses \cite{NielsenChuang}.

Here we focus on optical implementations of quantum information processing systems. Travelling optical pulses
are the natural candidates for the realization of quantum communication schemes and many experimental
demonstrations of quantum key distribution \cite{Gisin02,Grosshans03} and quantum teleportation schemes
~\cite{Bouwmeester97,Boschi98,Furusawa98,Bowen02,Zhang03} have been already performed. Optical systems have been
also proposed for the implementation of quantum computing, even though the absence of significant photon-photon
interactions is an obstacle for the realization of efficient two-qubit quantum gates, which are needed for
implementing universal quantum computation~\cite{NielsenChuang}. Various schemes have been proposed to
circumvent this problem. One is linear optics quantum computation~\cite{Knill01}, which is a probabilistic
scheme based on passive linear optical devices, efficient single photon sources and detectors and which
implicitly exploits the nonlinearity hidden in the photodetection process (see~\cite{OBrien03,Sanaka04} for some
preliminary demonstrations of this scheme). Other schemes explicitly exploit optical nonlinearities for quantum
gate implementations. Typical optical nonlinearities are too small to provide a substantial photon-photon
interaction, hence limiting the usefulness of an all-optical quantum gate. However there seems to be a way to
overcome the problem. Quantum interference effects associated with electromagnetically induced transparency
(EIT)~\cite{Boller91,Arimondo96,Harris97} have quite recently been shown to enhance these nonlinearities by as
much as 10 orders of magnitude~\cite{Hau99}. This enhancement is commonly exhibited by a weak probe beam in the
presence of another strong coupling beam when both impinge off-resonance on a three-level atomic sample at very
low temperatures.

The off-resonance condition is rather crucial to the observation of the enhancement and one can, in general,
identify two ways for attaining that. One is to introduce an additional laser beam whose detuning from a fourth
level is larger than the level linewidth~\cite{Schmidt96}. In this ``N'' configuration one of the ground levels
undergoes an ac-Stark shift which disturbs the EIT resonance condition and induces an effective Kerr
nonlinearity while keeping absorption negligible. Improvements by many orders of magnitude with respect to
conventional nonlinearities have indeed been observed in this way~\cite{Kang03}. In addition, strong
cross-phase modulation~\cite{Schmidt96} and photon blockade (i.e. strong self-phase modulation) have also been
predicted~\cite{Imamoglu97,Grangier98,Rebic99,Gheri99,Greentree00}.

Another and related way to obtain large nonlinearities consists in disturbing the exact two-photon resonance
condition in a $\Lambda$ configuration. This can be achieved by slightly mismatching the probe and coupling
field frequencies yet within the EIT transparency window making the dispersion of the probe field not exactly
zero. In this case enhanced Kerr nonlinearities have been observed in the $\Lambda$
configuration~\cite{Roch97,Wang01} and predicted in the so-called chain-$\Lambda$
configurations~\cite{Zubairy02,Matsko02,Matsko03,Greentree03,Ottaviani03}. By using this second approach
Ottaviani {\it et al.}~\cite{Ottaviani03} have shown that large cross-phase modulations that occur in an ``M''
configuration may lead to an all-optical two-qubit quantum phase gate (QPG) \cite{NielsenChuang,Lloyd95}, where
one qubit gets a phase shift dependent on the state of the other qubit. Here, the key element enabling large
cross-phase modulation is the possibility of group velocity matching. Large cross-phase modulations occur when
two optical pulses, a \emph{probe} and a \emph{trigger}, interact for a sufficiently long time. This happens
when their group velocities are both small and comparable~\cite{Lukin00,Lukin01} and there exists several ways
by which this can be done~\cite{Lukin00,Petrosyan02,Ottaviani03}.

This paper proposes an alternative scheme for phase gating that can greatly reduce, when compared with other
schemes, the experimental effort for its realizability. The mechanism relies on an enhanced cross-phase
modulation effect which occurs in a relatively simple and robust four atomic level \emph{tripod} configuration.
Our scheme only requires good control over frequencies and intensities of the laser beams. As in
Ref.~\cite{Ottaviani03}, we consider a QPG for qubits in which binary information is encoded in the polarization
of an optical field.


Optical QPG have been already experimentally studied. A conditional phase shift $\phi \simeq 16^{\circ}$ between
two frequency-distinct high-Q cavity modes, due to the effective cross modulation mediated by a beam of Cs
atoms, has been measured in Ref.~\cite{Turchette95}. However, the complete truth table of the gate has not been
determined in this experiment. A conditional phase shift $\phi \simeq 8^{\circ}$ has been instead obtained
between weak coherent pulses, using a second-order nonlinear crystal  \cite{Resch02}. However, this experiment
did not demonstrate a {\em bona fide} QPG because $\phi$ depends on the input states, and the gate can be
defined only for a restricted class of inputs (weak coherent states).

The four level tripod configuration that we adopt here has been extensively studied in the past few years. For
example, Unanyan {\it et al.}~\cite{Unanyan98} used a tripod configuration to achieve stimulated Raman adiabatic
passage (STIRAP) for creating an arbitrary coherent superposition of two atomic states in a controlled way.
Paspalakis {\em et al.}~\cite{Paspalakis02a,Paspalakis02b,Paspalakis02c}, in particular, developed the
interesting possibility of using a tripod scheme for efficient nonlinear frequency generation. Moreover, it was
shown that the group velocity of a probe pulse may be significantly reduced, as in conventional $\Lambda$
system~\cite{Paspalakis02a}. The work of Malakyan~\cite{Malakyan01} was the first to hint that the tripod scheme
may be used to entangle a pair of very weak optical fields in an atomic sample. This work has been recently
extended to the case of quantum probe and trigger fields in \cite{Petrosyan04}, where an adiabatic treatment
similar to that of \cite{Lukin00} is adopted.

The purpose of this paper is thus twofold. \textit{First}, we adopt a standard density matrix approach,
including spontaneous emission and dephasings, to analyse the nonlinear optical response of a four-level tripod
configuration. In particular, we examine the conditions under which large cross-Kerr nonlinearities may occur in
a cold atomic sample. \textit{Second}, we study the possibility of employing such an enhanced cross-phase
modulation to devise a polarization phase-gating mechanism which turns out to be rather robust and apt to actual
experimental investigations.

The paper is organized as follows. In Sec.~\ref{sec:dress}, dressed states of the atomic tripod are analyzed and
their significance emphasized. In Sec.~\ref{sec:bloch}, we solve the set of Bloch equations and derive
expressions for linear and nonlinear susceptibilities. In Sec.~\ref{sec:velocity} group velocity matching is
discussed in detail, while Sec.~\ref{sec:phasegate} discusses the operation of a polarization phase gate. We
summarize our results in Sec.~\ref{sec:conclusion}.

\section{Dressed States of the Tripod System \label{sec:dress}}

\begin{figure}[t]
  \includegraphics[scale=0.7]{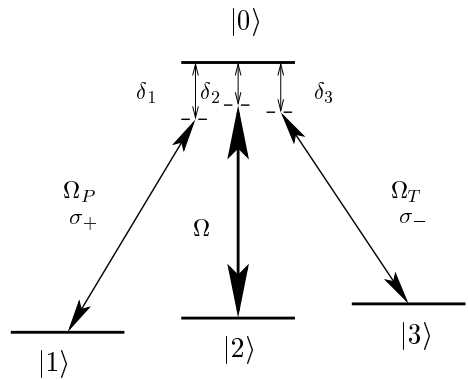}
  \caption{Energy level scheme for a tripod. Probe and trigger fields
have Rabi frequencies $\Omega_P$ and $\Omega_T$ and polarizations
$\sigma_+$ and $\sigma_-$. The pump Rabi frequency is $\Omega$ while
$\delta_j= \omega_0-\omega_j-\omega_j^{(L)}$ denote the laser
(frequency $\omega_j^{(L)}$) detunings from the respective
transitions $|j\rangle \leftrightarrow |0\rangle$.  \label{fig:tripod}}
\end{figure}

The energy level scheme of a tripod system is given in Fig.~\ref{fig:tripod}. Transitions $|1\rangle \rightarrow
|0\rangle$ and $|3\rangle \rightarrow |0\rangle$ are driven by a probe and trigger fields of respective Rabi
frequencies $\Omega_P$ and $\Omega_T$, while the transition $|2\rangle \rightarrow |0\rangle$ is driven by a
control (or pump) field of Rabi frequency $\Omega$. Moreover $\delta_j= \omega_0-\omega_j-\omega_j^{(L)}$ denote
the laser (frequency $\omega_j^{(L)}$) detunings from the respective transitions $|j\rangle \leftrightarrow
|0\rangle$. The system Hamiltonian, in the interaction picture and in the dipole and rotating wave
approximations, is given by
\begin{eqnarray}
   {\mathcal H}_{int} &=& \hbar\delta_1 \sigma_{00} +
\hbar(\delta_1-\delta_2) \sigma_{22} + \hbar(\delta_1-\delta_3)
\sigma_{33} \nonumber  \\
&\ &+ \hbar\left( \Omega_P^* \sigma_{10} + \Omega_P
\sigma_{01}\right) + \hbar\left( \Omega^* \sigma_{20} + \Omega
\sigma_{02}\right) \nonumber \\
&\ &+ \hbar\left( \Omega_T^* \sigma_{30} + \Omega_T
\sigma_{03}\right) ,\label{eq:Hint}
\end{eqnarray}
where $\sigma_{ij} = |i\rangle\langle j|$ are pseudospin atomic operators. Spontaneous emission and dephasing is
included below [see Eqs.~(\ref{eq:blocheqs})] in a standard way~\cite{Walls94}.

There are four eigenstates of
Hamiltonian~(\ref{eq:Hint})~\cite{Unanyan98}. When the three detunings are equal, $\delta_i = \delta$, $i=1,2,3$, two of them are
degenerate with energy equal to $\delta$ and assume the following form:
\begin{subequations}
\label{eq:darkstates}
\begin{eqnarray}
   |e_1\rangle &=& \frac{\Omega_T |1\rangle -
\Omega_P|3\rangle}{\sqrt{\Omega_P^2+\Omega_T^2}} , \\
   |e_2\rangle &=& \frac{\Omega_T \Omega_P |1\rangle + \Omega \Omega_P
|3\rangle - \left( \Omega_P^2 + \Omega_T^2 \right)
|2\rangle}{\sqrt{\left( \Omega_P^2+\Omega_T^2\right)\left(
\Omega_P^2+\Omega^2+\Omega_T^2 \right)}}.
\end{eqnarray}
\end{subequations}
Since these states do not contain any contribution of the excited state $|0\rangle$, they belong to the class of
dark states. The other two eigenstates have energies $\delta \pm \sqrt{\Omega_P^2+\Omega^2+\Omega_T^2}$ and are
\begin{equation}
   |e_\pm\rangle = \frac{\Omega_P |1\rangle \pm |0\rangle + \Omega_T
|3\rangle + \Omega |2\rangle}{\sqrt{\Omega_P^2+\Omega^2+\Omega_T^2}}
\label{eq:brightstates}.
\end{equation}
In the case of different detunings, the expression of the eigenstates becomes more complicated, and the
degeneracy of the two dark states is removed because their energies shift from $\delta$ to $\delta_2$ and
$\delta_3$ respectively.

The scope of the present paper is to show that the tripod configuration of Fig.~\ref{fig:tripod} allows to
achieve a giant cross-phase modulation between probe and trigger fields, based on the steep dispersion
associated with EIT. Necessary conditions for achieving a large cross-Kerr phase shift can be formulated as
follows: $(i)$ probe and trigger must be tuned to dark states, $(ii)$ the transparency frequency window for each
of these dark states has to be narrow and with a steep dispersion to enable significant group velocity
reduction, and $(iii)$ there must be a degree of symmetry between the two transparency windows so that trigger
and probe group velocities can be made to be equal~\cite{Lukin00,Petrosyan02,Ottaviani03}. These conditions can
be satisfied by taking all three detunings nearly equal. When the three detunings are equal the two dark states
are degenerate, giving a common transparency window for both fields. However, it can be seen that for perfectly
equal detunings the tripod system is linear, i.e., the dispersive nonlinearity vanishes [see
Eqs.~(\ref{eq:kerrsuscpandt})]. Hence, the exact resonance condition will have to be violated. We will show that
if the frequency mismatch is small (within the transparency window width), then strong, cross-Kerr modulation
with group velocity matching can still be achieved and phase gate operation realized.

\section{Bloch Equations and Susceptibilities \label{sec:bloch}}

The Bloch equations for the density matrix elements (including atomic spontaneous emission and dephasing) are
\begin{subequations}
\label{eq:blocheqs}
   \begin{eqnarray}
     i\dot{\rho}_{00} &=& -i(\gamma_{11}+\gamma_{22}+\gamma_{33})
\rho_{00} + \Omega_P^*\rho_{10} - \Omega_P\rho_{01} \nonumber \\
&\ &+ \Omega^*\rho_{20} - \Omega\rho_{02} + \Omega_T^*\rho_{30} -
\Omega_T\rho_{03}, \label{eq:bloch00} \\
     i\dot{\rho}_{11} &=& i\gamma_{11} \rho_{00} + i\gamma_{12}
\rho_{22} + i\gamma_{13} \rho_{33} + \Omega_P\rho_{01} -
\Omega_P^*\rho_{10} \label{eq:bloch11}, \\
     i\dot{\rho}_{22} &=& i\gamma_{22} \rho_{00} - i\gamma_{12}
\rho_{22} + i\gamma_{23} \rho_{33} + \Omega\rho_{02} -
\Omega^*\rho_{20} \label{eq:bloch22}, \\
     i\dot{\rho}_{33} &=& i\gamma_{33} \rho_{00} -
i(\gamma_{13}+\gamma_{23}) \rho_{33} + \Omega_T\rho_{03} -
\Omega_T^*\rho_{30} \label{eq:bloch33}, \\
     i\dot{\rho}_{10} &=& -\Delta_{10} \rho_{10} + \Omega_P \rho_{00}
- \Omega_P \rho_{11} - \Omega\rho_{12} - \Omega_T\rho_{13}
\label{eq:bloch10}, \\
     i\dot{\rho}_{20} &=& -\Delta_{20} \rho_{20} + \Omega \rho_{00} -
\Omega \rho_{22} - \Omega_P\rho_{21} - \Omega_T\rho_{23}
\label{eq:bloch20}, \\
     i\dot{\rho}_{30} &=& -\Delta_{30} \rho_{30} + \Omega_T \rho_{00}
- \Omega_P \rho_{33} - \Omega_P\rho_{31} - \Omega\rho_{32}
\label{eq:bloch30}, \\
     i\dot{\rho}_{12} &=& -\Delta_{12} \rho_{12} + \Omega_P\rho_{02} -
\Omega^*\rho_{10} \label{eq:bloch12}, \\
     i\dot{\rho}_{13} &=& -\Delta_{13} \rho_{13} + \Omega_P\rho_{03} -
\Omega_T^*\rho_{10} \label{eq:bloch13}, \\
     i\dot{\rho}_{23} &=& -\Delta_{23} \rho_{23} + \Omega\rho_{03} -
\Omega_T^*\rho_{20} \label{eq:bloch23},
   \end{eqnarray}
\end{subequations}
where $\rho_{ij} = {\rm Tr}\{\sigma_{ji}\rho\} = \langle
i|\rho|j\rangle$. Decay rates $\gamma_{ij}$ describe decay of
populations and coherences, $\Delta_{j0} = \delta_j +i\gamma_{j0}$
and $\Delta_{ij} = \delta_j - \delta_i - i\gamma_{ij}$, with $i,j =
1,2,3$.

We consider the steady state solutions to the Bloch equations. When the intensity of the pump field is stronger
than the intensity of both probe and trigger $|\Omega|^2 \gg |\Omega_{P,T}|^2$, and the detunings and decay
rates are of the same order of magnitude, the final population distribution will be symmetric with respect to
the $1 \leftrightarrow 3$ exchange, i.e., $\rho_{11} \approx \rho_{33} \approx 1/2$, with the population of the
other two levels vanishing. This allows to decouple the equations for the populations from those of the
coherences and to obtain the steady state solution for the latter, yielding the probe and trigger
susceptibilities according to
\begin{subequations}
   \begin{eqnarray}
     \chi_P &=& -\lim_{t\rightarrow\infty}\,
\frac{\mathcal{N}|\bm{\mu}_P|^2}{\hbar\epsilon_0} \times
\frac{\rho_{10}(t)}{\Omega_P} , \\
     \chi_T &=& -\lim_{t\rightarrow\infty}\,
\frac{\mathcal{N}|\bm{\mu}_T|^2}{\hbar\epsilon_0} \times
\frac{\rho_{30}(t)}{\Omega_T} ,
   \end{eqnarray}
\end{subequations}
where $\mathcal{N}$ is the atomic density and $\bm{\mu}_{P,T}$ the electric dipole matrix elements for probe and
trigger transitions respectively. Rabi frequencies are defined in terms of electric field amplitudes $E_{P,T}$
as $\Omega_{P,T} = -\left(\bm{\mu}_{P,T} \cdot \bm{\varepsilon}_{P,T}\right) E_{P,T}/\hbar$, with
$\bm{\varepsilon}_{P,T}$ being the polarization unit vector of probe and trigger beams. The resulting general
expression for the steady-state $(ss)$ probe and trigger susceptibilities are obtained from
\begin{subequations}
\label{eq:bigsuscgen}
\begin{eqnarray}
\frac{\left(\rho_{10}\right)_{ss}}{\Omega_P} &=& \left( 1+\frac{1}{4}
\frac{\left(\Delta_{12}\Delta_{23}/\Delta_{13}^2\right)|\Omega_P|^2|\Omega_T|^2}{\left(\Delta_{10}\Delta_{12}-
|\Omega|^2\right)\left(\Delta_{30}^*\Delta_{23}-|\Omega|^2\right)}\right)^{-1} \label{eq:rho10}
\nonumber \\
&\times&
\left\{-\frac{1}{2}\frac{\Delta_{12}\Delta_{13}}{\Delta_{10}\Delta_{12}\Delta_{13}-\Delta_{13}|\Omega|^2-\Delta_{12}|\Omega_T|^2}\right.
\nonumber \\
&\ &\left.
-\frac{1}{2}\frac{\Delta_{12}\Delta_{13}\Delta_{23}|\Omega_T|^2}{\Delta_{30}^*\Delta_{13}\Delta_{23}-\Delta_{13}|\Omega|^2-\Delta_{23}|\Omega_P|^2}
\right\}, \\
\frac{\left(\rho_{30}\right)_{ss}}{\Omega_T} &=& \left( 1+\frac{1}{4}
\frac{\left(\Delta_{23}^*\Delta_{12}^*/\Delta_{13}^{*
2}\right)|\Omega_P|^2|\Omega_T|^2}{\left(\Delta_{30}\Delta_{23}^*-|\Omega|^2\right)\left(\Delta_{10}^*\Delta_{12}^*-|\Omega|^2\right)}\right)^{-1}
\nonumber \\
&\times&
\left\{-\frac{1}{2}\frac{\Delta_{23}^*\Delta_{13}^*}{\Delta_{30}\Delta_{23}^*\Delta_{13}^*-\Delta_{13}^*|\Omega|^2-\Delta_{23}^*|\Omega_P|^2}\right.
\nonumber \\
&\ &\left.
-\frac{1}{2}\frac{\Delta_{23}^*\Delta_{13}^*\Delta_{12}^*|\Omega_P|^2}{\Delta_{10}^*\Delta_{13}^*\Delta_{12}^*-\Delta_{13}^*|\Omega|^2-\Delta_{12}^*|\Omega_T|^2}
\right\}.
\end{eqnarray}
\end{subequations}
We are interested in the cross-phase modulation between the probe and trigger fields. Therefore, we keep the two
lowest order contributions in trigger and probe: linear and third-order nonlinear susceptibilities, while
neglecting the higher orders in the expansion. This yields
\begin{subequations}
\label{eq:suscgen}
\begin{eqnarray}
   \chi_P &=& \chi_P^{(1)} + \chi_P^{(3)} |E_T|^2, \\
   \chi_T &=& \chi_T^{(1)} + \chi_T^{(3)} |E_P|^2
\end{eqnarray}
\end{subequations}
that is, each susceptibility has a linear and a cross--Kerr nonlinear term, while self-phase modulation terms
are of higher order. Both susceptibilities have a linear contribution because of the nonzero stationary
population in levels $1$ and $3$. Linear susceptibilities are given by
\begin{subequations}
\label{eq:linsuscpandt}
\begin{eqnarray}
   \chi_P^{(1)} &=& \frac{\mathcal{N}|\bm{\mu}_P|^2}{\hbar\epsilon_0}
\times
\frac{1}{2}\frac{\Delta_{12}}{\Delta_{10}\Delta_{12}-|\Omega|^2},
\label{eq:linsuscp} \\
   \chi_T^{(1)} &=& \frac{\mathcal{N}|\bm{\mu}_T|^2}{\hbar\epsilon_0}
\times
\frac{1}{2}\frac{\Delta_{23}^*}{\Delta_{30}\Delta_{23}^*-|\Omega|^2},
\label{eq:linsusct}
\end{eqnarray}
\end{subequations}
where the factor $1/2$ in each equation comes from the symmetric steady state population distribution. The
cross-Kerr susceptibilities are instead given by
\begin{subequations}
\label{eq:kerrsuscpandt}
\begin{eqnarray}
   \chi_P^{(3)} &=& \mathcal{N}\,
\frac{|\bm{\mu}_P|^2|\bm{\mu}_T|^2}{\hbar^3\epsilon_0} \times
\frac{1}{2}\frac{\Delta_{12}/\Delta_{13}}{\Delta_{10}\Delta_{12}-|\Omega|^2}
\nonumber \\
&\ & \times \left(
\frac{\Delta_{12}}{\Delta_{10}\Delta_{12}-|\Omega|^2}+\frac{\Delta_{23}}{\Delta_{30}^*\Delta_{23}-|\Omega|^2}
\right), \label{eq:kerrsuscp} \\
   \chi_T^{(3)} &=& \mathcal{N}\,
\frac{|\bm{\mu}_T|^2|\bm{\mu}_P|^2}{\hbar^3\epsilon_0} \times
\frac{1}{2}\frac{\Delta_{23}^*/\Delta_{13}^*}{\Delta_{30}\Delta_{23}^*-|\Omega|^2}
\nonumber \\
&\ &\times \left(
\frac{\Delta_{12}^*}{\Delta_{10}^*\Delta_{12}^*-|\Omega|^2}+\frac{\Delta_{23}^*}{\Delta_{30}\Delta_{23}^*-|\Omega|^2}
\right) . \label{eq:kerrsusct}
\end{eqnarray}
\end{subequations}
Note that Eqs.~(\ref{eq:linsuscpandt}) and also Eqs.~(\ref{eq:kerrsuscpandt}) are completely symmetric with
respect to the $1 \leftrightarrow 3$ exchange~\footnote{The full symmetry also requires $|\bm{\mu}_T|^2 =
|\bm{\mu}_P|^2$, which is fulfilled for the proposed $^{87}$Rb scheme, see Sec.~\ref{sec:phasegate}.}. This
exchange symmetry is ensured by the complex conjugate terms in~(\ref{eq:linsusct}) and~(\ref{eq:kerrsusct}) and
it is expected because of the symmetry of the population distribution. Note also that in the absence of
dephasing, the nonlinear susceptibility has a singularity at $\delta_1 = \delta_3$. The necessary regularization
is provided by a nonzero dephasing term $i\gamma_{13}$.

Paspalakis and Knight~\cite{Paspalakis02a} have recently analyzed the properties of the tripod system in a
somewhat different setup. It is nevertheless instructive to compare the results of this Section with theirs. In
the scheme of~\cite{Paspalakis02a}, population is assumed to be initially in the ground state $|1\rangle$.
Provided that $|\Omega_P|^2 \ll |\Omega|^2,\, |\Omega_T|^2$ population remains in $|1\rangle$ in the steady
state. Paspalakis and Knight calculate the expression for probe susceptibility to the first order in $\Omega_P$.
It is easy to see that their expression is consistent (up to a factor $1/2$ determined by the different
population distribution) to our result in Eq.~(\ref{eq:rho10}): considering only terms to the first order in
$\Omega_P$ leaves only the first term in the curly brackets of~(\ref{eq:rho10}). Additional terms in
Eqs.~(\ref{eq:bigsuscgen}) arise because we are looking for a cross-Kerr nonlinearity in both probe and trigger,
so that all the terms of third order have to be included.

\section{Group Velocity Matching \label{sec:velocity}}

\begin{figure}[t]
   \includegraphics[scale=0.75]{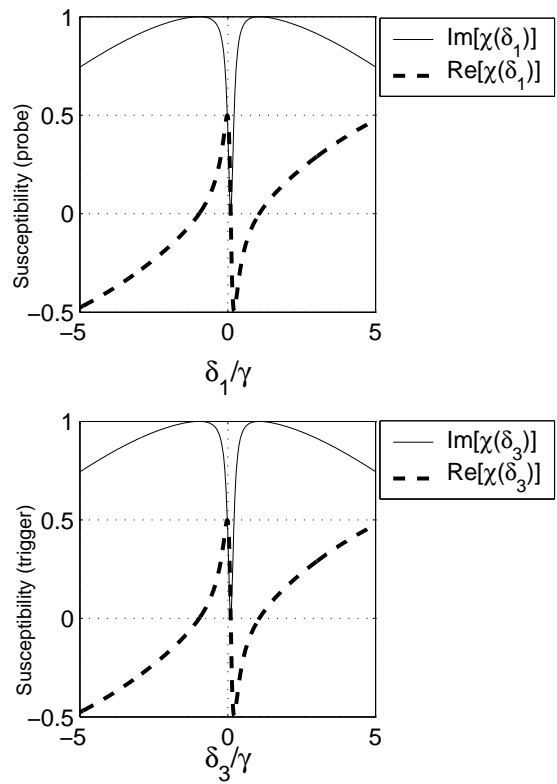}
   \caption{Probe absorption and dispersion $\chi(\delta_1) = \hbar\epsilon_0\chi_P/(\mathcal{N}|\bm{\mu}_P|^2) = \left( \frac{\rho_{10}}{\Omega_P}\right)_{ss}\ [\gamma^{-1}]$ (\textit{upper frame}) vs. the probe detuning $\delta_1/\gamma$ when
$\delta_3 = 0.1\gamma$ and $\delta_2 = 0.1\gamma$. Trigger absorption and dispersion $\chi(\delta_3) =
\hbar\epsilon_0\chi_T/(\mathcal{N}|\bm{\mu}_T|^2) = \left(\frac{\rho_{30}}{\Omega_T}\right)_{ss}\ [\gamma^{-1}]$
(\textit{lower frame}) vs. the trigger detuning $\delta_3/\gamma$ when $\delta_1 = 0.1\gamma$ and $\delta_2 =
0.1\gamma$. In both cases we take the Rabi frequencies as $\Omega_P = \Omega_T = 0.1\gamma$, $\Omega = \gamma$.
\label{fig:symtrans}}
\end{figure}

The linear and nonlinear susceptibilities of Eqs.~(\ref{eq:linsuscpandt}) and (\ref{eq:kerrsuscpandt}) have all
the properties required for a large cross-phase modulation. In fact, our tripod system can be seen as formed by
two adjacent $\Lambda$ systems, one involving the probe field and one involving the trigger field, sharing the
same control field. Therefore both fields exhibit EIT, which here manifests itself through the presence of two
generally distinct transparency windows, corresponding to the two dark states of Eq.~(\ref{eq:darkstates}).
Perfect EIT for both fields takes place when the two transparency windows coincide, i.e., when the two dark
states are degenerate, which is achieved when the three detunings $\delta_i$ are all equal. In this case, all
physical effects related to standard EIT are present and in particular the steep dispersion responsible for the
reduction of the group velocity which is at the basis of the giant cross-Kerr nonlinearity (see
Fig.~\ref{fig:symtrans}). The condition of equal detunings (exact double EIT-resonance condition) is important
also for another reason. In fact, together with the symmetry of Eqs.~(\ref{eq:linsuscpandt}) and
(\ref{eq:kerrsuscpandt}) with respect to the $1 \leftrightarrow 3$ exchange, it also guarantees identical
dispersive properties for probe and trigger and therefore the same group velocity. As first underlined by Lukin
and Imamo\u{g}lu~\cite{Lukin00}, group velocity matching is another fundamental condition for achieving a large
nonlinear mutual phase shift because only in this way the two optical pulses interact in a transparent nonlinear
medium for a sufficiently long time.

The group velocity of a light pulse is given in general by $v_g = c/(1+n_g)$, where $c$ is the speed of light in
vacuum and
\begin{equation}
n_g = \frac{1}{2} {\rm Re}[\chi] + \frac{\omega_0}{2} \left(\frac{\partial
{\rm Re}[\chi]}{\partial \omega} \right)_{\omega_0} \label{eq:ng}
\end{equation}
is the group index, $\omega_0$ being the laser frequency. The group index of Eq.~(\ref{eq:ng}) is essentially
determined by the linear susceptibility $\chi^{(1)}$, because contributions from the nonlinear terms are orders
of magnitude smaller and can be neglected. Using Eqs.~(\ref{eq:linsuscpandt}), it is possible to get a simple
expression for the two group velocities in the case of equal detunings. This condition corresponds to the center
of the transparency window for each field, where Re[$\chi^{(1)}$] vanishes, and the group velocity is reduced
due to a large dispersion gradient. One has
\begin{subequations}
\label{eq:vg}
\begin{eqnarray}
   \left( v_g \right)_P &\approx& \frac{4\hbar c \epsilon_0}{\omega_P
\mathcal{N} |\bm{\mu}_P|^2} \left( |\Omega|^2 + |\Omega_T|^2\right) ,
\\
   \left( v_g \right)_T &\approx& \frac{4\hbar c \epsilon_0}{\omega_T
\mathcal{N} |\bm{\mu}_T|^2} \left( |\Omega_P|^2 +|\Omega|^2 \right),
\end{eqnarray}
\end{subequations}
so that, as expected from the $1 \leftrightarrow 3$ symmetry, group velocity matching is achieved
for $|\Omega_P| = |\Omega_T|$.

Unfortunately, it is possible to check from Eqs.~(\ref{eq:kerrsuscpandt}) that when $\delta_i = \delta$,
$\forall i$ exactly, the system becomes linear, i.e., the real part of the nonlinear susceptibilities vanish and
there is no cross-phase modulation. This means that we have to ``disturb'' the exact EIT resonance conditions,
by taking slightly different detunings. This is a general conclusion, valid for any atomic level scheme
resembling multiple $\Lambda$ systems~\cite{Ottaviani03,Matsko02,Matsko03,Greentree03}. If the double
EIT-resonance condition is disturbed by a small amount, one remains within the common transparency window and
the absorption is still negligible. Moreover, the two group velocities can be matched also in the non-resonant
case. In fact, from the symmetry of Eqs.~(\ref{eq:linsuscpandt}), one has that the gradients - and hence the
group velocities - can be kept symmetric and all the conclusions for the exact resonance remain valid in the
vicinity of resonance as well.

\section{Phase Gate Operation \label{sec:phasegate}}

A medium able to realize a significant cross-phase modulation is the key ingredient for the implementation of a
quantum gate between two optical qubits. Such a gate requires the existence of conditional quantum dynamics,
which is realized in the cross-Kerr effect where an optical field acquires a phase shift conditional to the
state of another optical field. Using cross phase modulation one can implement a QPG, defined by the following
input-output relations $|i\rangle _{1}|j\rangle _{2} \rightarrow \exp\left\{i \phi_{ij} \right\}|i\rangle
_{1}|j\rangle _{2} $, with $i,j=0,1$ denoting the logical qubit bases. This gate is a universal two-qubit gate,
that is, it is able to entangle two initially factorized qubits, when the conditional phase shift
$\phi=\phi_{11}+\phi_{00}-\phi_{10}-\phi_{01}\neq 0$ \cite{Turchette95,Lloyd95,NielsenChuang}. A natural choice
for encoding binary information in optical beams is to use the polarization degree of freedom, where the two
logical basis states $|0\rangle $ and $|1\rangle $ correspond to two orthogonal polarizations. In this case one
can implement a universal QPG if a nontrivial cross-phase modulation between probe and trigger fields arise for
only one of the four possible input configurations of their polarization.


A possible experimental configuration employing the giant Kerr nonlinear phase shift achievable in the tripod
scheme discussed above is provided by a $^{87}$Rb atoms confined in a magneto-optical trap (MOT), in which
states $|1\rangle$, $|2\rangle$ and $|3\rangle$ correspond to the ground state Zeeman sublevels $|5S_{1/2}, F =
1, m = \{ -1, 0, 1 \} \rangle$, and state $|0\rangle$ corresponds to the excited state $|5P_{3/2}, F = 0
\rangle$. One realizes the tripod scheme of Fig.~1 (and therefore a significant nonlinear phase shift) only when
the probe has $\sigma_+$ polarization and the trigger has $\sigma_-$ polarization. When either the probe or the
trigger polarizations (or both) are changed, the phase shifts acquired by the two pulses do not involve the
nonlinear susceptibilities and are different, so that the resulting conditional phase shift is nonzero. In fact,
when they have the ``wrong'' polarization (probe $\sigma^{-}$ polarized or trigger $\sigma^{+}$ polarized) there
is no sufficiently close level which the atoms can be driven to and the fields acquire the trivial vacuum phase
shift $\phi_{0}^{j}=k_{j}l$, $j=P,T$, where $l$ in the length of the medium. Instead, when only one of them has
the right polarization, it acquires a linear phase shift $\phi_{lin}^{j}$, $j=P,T$, where
\begin{equation}
\label{eq:linphase}
   \phi_{lin}^j = k_jl \left( 1 + 2\pi \chi_j^{(1)} \right).
\end{equation}
Denoting with $\phi_{nlin}^{P,T}$ the corresponding probe and trigger nonlinear phase shift when the tripod
configuration is realized, we arrive at the following truth table for the polarization QPG
\begin{subequations}
\label{eq:QPGtruthtable}
\begin{eqnarray}
   |\sigma^-\rangle_P |\sigma^-\rangle_T &\rightarrow&
e^{-i(\phi_0^P+\phi_{lin}^T)}|\sigma^-\rangle_P |\sigma^-\rangle_T, \\
   |\sigma^-\rangle_P |\sigma^+\rangle_T &\rightarrow&
e^{-i(\phi_0^P+\phi_0^T)}|\sigma^-\rangle_P |\sigma^+\rangle_T, \\
   |\sigma^+\rangle_P |\sigma^+\rangle_T &\rightarrow&
e^{-i(\phi_{lin}^P+\phi_0^T)}|\sigma^+\rangle_P |\sigma^+\rangle_T, \\
   |\sigma^+\rangle_P |\sigma^-\rangle_T &\rightarrow&
e^{-i(\phi_+^P+\phi_-^T)}|\sigma^+\rangle_P |\sigma^-\rangle_T,
\end{eqnarray}
\end{subequations}
with the conditional phase shift being
\begin{equation}
\phi = \phi_+^P + \phi_-^T - \phi_{lin}^P - \phi_{lin}^T,
\end{equation}
with $\phi_+^P = \phi_{lin}^P + \phi_{nlin}^P$ and $\phi_-^T = \phi_{lin}^T + \phi_{nlin}^T$. Notice that only
the nonlinear part contributes to the conditional phase shift. The truth table of Eqs.~(\ref{eq:QPGtruthtable})
differs from that of Ottaviani {\it et al.}~\cite{Ottaviani03} only in the presence of an additional linear
phase shift for the trigger field, which is a consequence of the fact that also level $3$ is populated by one
half of the atoms.

For a Gaussian trigger pulse of time duration $\tau_T$, whose peak Rabi frequency is $\Omega_T$, moving with
group velocity $v_g^T$ through the atomic sample, the nonlinear probe phase shift can be written as
\begin{subequations}
\begin{equation}
   \phi_{nlin}^P = k_Pl
\frac{\pi^{3/2}\hbar^2|\Omega_T|^2}{4|\bm{\mu}_T|^2}\,
\frac{\rm{erf}[\zeta_P]}{\zeta_P} \, {\rm Re}[\chi_P^{(3)}],
\end{equation}
where $\zeta_P = (1-v_g^P/v_g^T)\sqrt{2}l/v_g^P\tau_T$. The trigger phase shift is simply obtained by changing
$P \leftrightarrow T$ in the equation above
\begin{equation}
   \phi_{nlin}^T = k_Tl
\frac{\pi^{3/2}\hbar^2|\Omega_P|^2}{4|\bm{\mu}_P|^2}\,
\frac{\rm{erf}[\zeta_T]}{\zeta_T} \, {\rm Re}[\chi_T^{(3)}],
\end{equation}
\end{subequations}
with the same appropriate changes in the definition of $\zeta_T$.

In the $^{87}$Rb level configuration chosen above, the decay rates are equal $\gamma_{j0}= \gamma$, and we
choose equal dephasing rates $\gamma_{ij}= \gamma_d$ for simplicity. For $\Omega_P \approx \Omega_T = 0.1
\gamma$, $\Omega = \gamma$, and detunings $\delta_1 = 20.01\gamma$, $\delta_2 = 20\gamma$, $\delta_3 =
20.02\gamma$, by assuming a low dephasing rate $\gamma_d = 10^{-2} \gamma$, we obtain a conditional phase shift
of $\pi$ radians, over the interaction length $l = 1.6$ mm, density $\mathcal{N} = 3 \times 10^{13}$ cm$^{-3}$.
With these parameters, group velocities are virtually the same, giving $\rm{erf}[\zeta_P]/\zeta_P =
\rm{erf}[\zeta_T]/\zeta_T \approx 2/\sqrt{\pi}$. These parameters correspond to a case where a polarization
qubits are encoded into a single photon wave packets, a desired setup for the implementation of a QPG operation.
As discussed in Ref.~\cite{Ottaviani03}, the proposed QPG can also be demonstrated by using post selection of
single photon coherent pulses instead of single photon wave packets~\footnote{Realization of QPG operation for a
single photons relies on an assumption of negligible transfer of fluctuations from a classical pump to probe and
trigger single photon pulses/wave packets. Evaluation of the validity of this assumption will be given in our
forthcoming publication.}.

\begin{figure}[t]
   \includegraphics[scale=0.75]{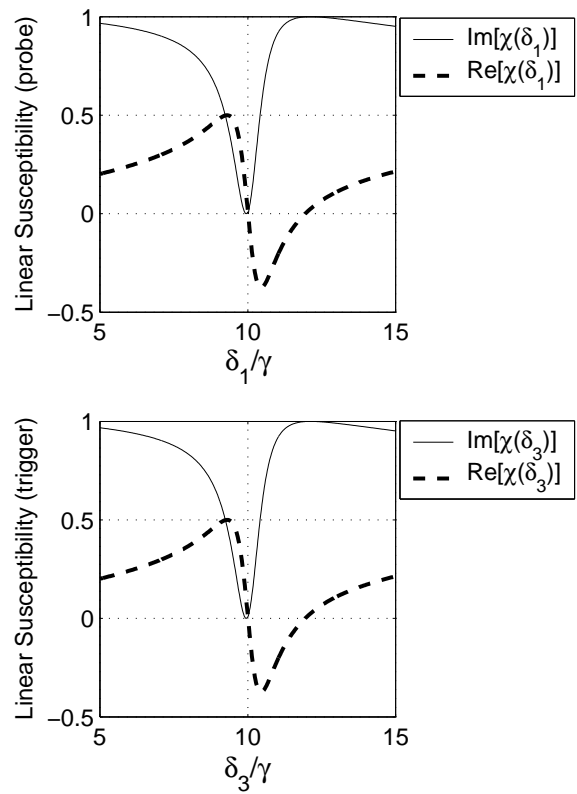}
   \caption{Probe absorption and dispersion $\chi(\delta_1) = \hbar\epsilon_0\chi_P/(\mathcal{N}|\bm{\mu}_P|^2) = \left( \frac{\rho_{10}}{\Omega_P}\right)_{ss}\ [\gamma^{-1}]$ (\textit{upper frame}) vs. the probe detuning $\delta_1/\gamma$ when
$\delta_3 = 10.02\gamma$ and $\delta_2 = 10\gamma$. Trigger absorption and dispersion $\chi(\delta_3) =
\hbar\epsilon_0\chi_T/(\mathcal{N}|\bm{\mu}_T|^2)\left( \frac{\rho_{30}}{\Omega_T}\right)_{ss}\ [\gamma^{-1}]$
(\textit{lower frame}) vs. the trigger detuning $\delta_3/\gamma$ when $\delta_1 = 10.01\gamma$ and $\delta_2 =
10\gamma$. In both cases we take the Rabi frequencies as $\Omega_P = \Omega_T = \gamma$, $\Omega = 4.5\gamma$.
\label{fig:symsusc}}
\end{figure}

Strong cross-phase modulation can also be achieved with classical fields, and we propose here alternative set of
parameters that can be used to achieve this. For (classical) Rabi frequencies $\Omega_P \approx \Omega_T =
\gamma$, $\Omega = 4.5 \gamma$, and detunings $\delta_1 = 10.01\gamma$, $\delta_2 = 10\gamma$, $\delta_3 =
10.02\gamma$, a conditional phase shift of $\pi$ radians, over the interaction length $l = 0.7$ cm, density
$\mathcal{N} = 3 \times 10^{12}$ cm$^{-3}$ is obtained. Again, with these parameters, group velocities are the
same. Probe and trigger susceptibilities corresponding to these parameter values are shown in
Fig.~\ref{fig:symsusc}. The above parameters are chosen to correspond to those obtained with cold atoms in a
MOT. Alternatively, a gas cell of standard length between 2.5 cm and 10 cm can be considered, but the increase
in length is then compensated with a lower density. This shows that a demonstration of a deterministic
polarization QPG can be made using present technologies.

As discussed above, we had to move from the exact double EIT-resonance condition in order to have a nonzero
nonlinearity and in such a condition the linear susceptibilities do not vanish. Actually, the linear
contribution is predominant. In fact, the ratios of nonlinear to linear phase shifts are given by
\begin{subequations}
\label{eq:phaseratios}
\begin{eqnarray}
 &&  \frac{\phi_P^{nlin}}{\phi_P^{lin}} = \frac{|\Omega_T|^2}{4} \nonumber \\
&& \times {\rm Re} \left[ \frac{1}{\Delta_{13}}\left(
\frac{\Delta_{12}}{\Delta_{10}\Delta_{12}-|\Omega|^2} +
\frac{\Delta_{23}}{\Delta^*_{30}\Delta_{23}-|\Omega|^2} \right)
\right] , \\
&&   \frac{\phi_T^{nlin}}{\phi_T^{lin}} = \frac{|\Omega_P|^2}{4} \nonumber \\
&& \times {\rm Re} \left[ \frac{1}{\Delta_{13}^*}\left(
\frac{\Delta_{12}}{\Delta_{10}^*\Delta_{12}^*-|\Omega|^2} +
\frac{\Delta_{23}^*}{\Delta^*_{30}\Delta_{23}^*-|\Omega|^2} \right)
\right] ,
\end{eqnarray}
\end{subequations}
and for the above choice of parameters, they are of order $\sim 1/43$ for the first (quantum) set of parameters
and $\sim 1/64$ for the second (semiclassical) set of parameters. This means that under the optimal conditions
corresponding to a $\pi$ conditional phase shift, the total phase shift in each input--output transformation is
very large, of the order of $45 \pi$ and $65 \pi$, respectively. The experimental demonstration of the QPG
requires the measurement of the conditional phase shift, i.e., of a phase difference and therefore it is
important to keep the errors in the phase measurements small. These errors are mainly due to the fluctuations of
the laser intensities and of the detunings. In particular, intensity fluctuations of 1\% yield an error of about
4\% in the phase measurement. It is more important to minimize the effects of relative detuning fluctuations but
this can be achieved by taking all lasers tightly phase locked to each other.

\begin{figure}[t]
  \includegraphics[width=0.45\textwidth,height=0.4\textwidth]{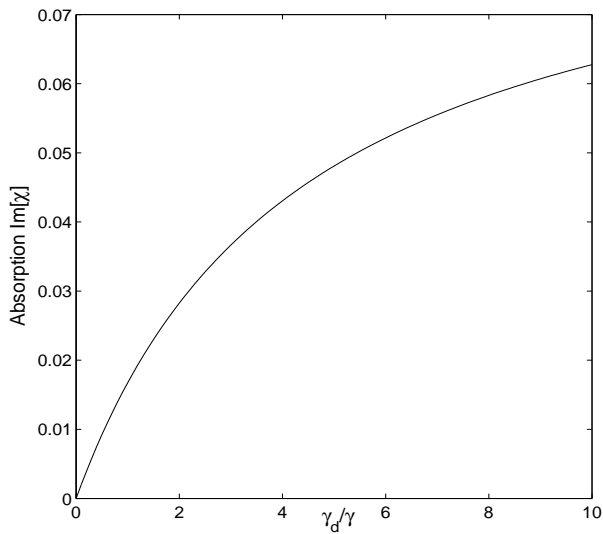}
  \caption{Probe absorption (scaled) at the center of probe transparency window, plotted against the dephasing rate, for
$\Omega_P = \Omega_T = \gamma$, $\Omega = 4.5\gamma$, $\delta_j = 0$.
\label{fig:abs}}
\end{figure}

Another important limitation is that due to dephasing of the ground state coherences, whose main effect is to
increase absorption. When the polarization qubits are carried by classical pulses one has only to be sure that
absorption is not too large, i.e., that it does not dominate over the nonlinear dispersion. Absorption is
instead a more crucial issue in the case of single photon polarization qubits. In fact a non negligible
absorption implies a nonzero gate failure probability (one or both qubits missing at the output), making
therefore the present QPG, which is deterministic in principle, a probabilistic gate. In our scheme, it can be
checked that, if the dephasings do not become very large, i.e., $\gamma_d = 2\pi\ \times$ 10 kHz, or $\gamma_d
\sim 10^{-2}\gamma$, this increase of absorption is negligible, as shown in Fig.~\ref{fig:abs}.

It should be mentioned that the conclusion above holds for strong control field strengths of order $\Omega \sim
\gamma$. If a weaker control is used, the dephasing must also be lower in order to keep absorption negligible.

\section{Conclusion \label{sec:conclusion}}

In this paper we have studied the nonlinear response of a four-level atomic sample in a tripod configuration to
an incident probe and trigger field. The resulting large cross-Kerr modulation between probe and trigger enables
one to implement a phase gate with a conditional phase shift of the order of $\pi$. The main advantage of our
proposal lies in its experimental feasibility which has been assessed through a detailed study of the
requirements needed to observe such a large shift in a cold atomic gas of $^{87}$Rb atoms in a MOT.

\begin{acknowledgments}
We acknowledge enlightening discussions with P. Grangier, F. T. Arecchi and M. Inguscio. We greatly acknowledge
support from the MURST (\emph{Actione Integrada Italia-Spagna}), the MIUR (PRIN 2001 \emph{Quantum
Communications with Slow Light}) and by MCyT and FEDER (\emph{project BFM2002-04369-C04-02}).
\end{acknowledgments}


\begin{thebibliography}{99}
\expandafter\ifx\csname natexlab\endcsname\relax\def\natexlab#1{#1}\fi \expandafter\ifx\csname
bibnamefont\endcsname\relax
  \def\bibnamefont#1{#1}\fi
\expandafter\ifx\csname bibfnamefont\endcsname\relax
  \def\bibfnamefont#1{#1}\fi
\expandafter\ifx\csname citenamefont\endcsname\relax
  \def\citenamefont#1{#1}\fi
\expandafter\ifx\csname url\endcsname\relax
  \def\url#1{\texttt{#1}}\fi
\expandafter\ifx\csname urlprefix\endcsname\relax\def\urlprefix{URL }\fi \providecommand{\bibinfo}[2]{#2}
\providecommand{\eprint}[2][]{\url{#2}}

\bibitem[{\citenamefont{Nielsen and Chuang}(2000)}]{NielsenChuang}
\bibinfo{author}{\bibfnamefont{M.~A.} \bibnamefont{Nielsen}} \bibnamefont{and}
  \bibinfo{author}{\bibfnamefont{I.~L.} \bibnamefont{Chuang}},
  \emph{\bibinfo{title}{Quantum Computation and Quantum Information}}
  (\bibinfo{publisher}{Cambridge University Press},
  \bibinfo{address}{Cambridge}, \bibinfo{year}{2000}).

\bibitem[{\citenamefont{Gisin et~al.}(2002)\citenamefont{Gisin, Ribordy,
  Tittel, and Zbinden}}]{Gisin02}
\bibinfo{author}{\bibfnamefont{N.}~\bibnamefont{Gisin}},
  \bibinfo{author}{\bibfnamefont{G.}~\bibnamefont{Ribordy}},
  \bibinfo{author}{\bibfnamefont{W.}~\bibnamefont{Tittel}}, \bibnamefont{and}
  \bibinfo{author}{\bibfnamefont{H.}~\bibnamefont{Zbinden}},
  \bibinfo{journal}{Rev. Mod. Phys.} \textbf{\bibinfo{volume}{74}},
  \bibinfo{pages}{145} (\bibinfo{year}{2002}).

\bibitem[{\citenamefont{Grosshans et~al.}(2003)\citenamefont{Grosshans, Assche,
  Wenger, Brouri, Cerf, and Grangier}}]{Grosshans03}
\bibinfo{author}{\bibfnamefont{F.}~\bibnamefont{Grosshans}},
  \bibinfo{author}{\bibfnamefont{G.~V.} \bibnamefont{Assche}},
  \bibinfo{author}{\bibfnamefont{J.}~\bibnamefont{Wenger}},
  \bibinfo{author}{\bibfnamefont{R.}~\bibnamefont{Brouri}},
  \bibinfo{author}{\bibfnamefont{N.~J.} \bibnamefont{Cerf}}, \bibnamefont{and}
  \bibinfo{author}{\bibfnamefont{P.}~\bibnamefont{Grangier}},
  \bibinfo{journal}{Nature} \textbf{\bibinfo{volume}{421}}, \bibinfo{pages}{238
  } (\bibinfo{year}{2003}).

\bibitem[{\citenamefont{Bouwmeester et~al.}(1997)\citenamefont{Bouwmeester,
  Pan, Mattle, Eibl, Weinfurter, and Zeilinger}}]{Bouwmeester97}
\bibinfo{author}{\bibfnamefont{D.}~\bibnamefont{Bouwmeester}},
  \bibinfo{author}{\bibfnamefont{J.-W.} \bibnamefont{Pan}},
  \bibinfo{author}{\bibfnamefont{K.}~\bibnamefont{Mattle}},
  \bibinfo{author}{\bibfnamefont{M.}~\bibnamefont{Eibl}},
  \bibinfo{author}{\bibfnamefont{H.}~\bibnamefont{Weinfurter}},
  \bibnamefont{and}
  \bibinfo{author}{\bibfnamefont{A.}~\bibnamefont{Zeilinger}},
  \bibinfo{journal}{Nature} \textbf{\bibinfo{volume}{390}},
  \bibinfo{pages}{575} (\bibinfo{year}{1997}).

\bibitem[{\citenamefont{Boschi et~al.}(1998)\citenamefont{Boschi, Branca,
  DeMartini, Hardy, and Popescu}}]{Boschi98}
\bibinfo{author}{\bibfnamefont{D.}~\bibnamefont{Boschi}},
  \bibinfo{author}{\bibfnamefont{S.}~\bibnamefont{Branca}},
  \bibinfo{author}{\bibfnamefont{F.}~\bibnamefont{DeMartini}},
  \bibinfo{author}{\bibfnamefont{L.}~\bibnamefont{Hardy}}, \bibnamefont{and}
  \bibinfo{author}{\bibfnamefont{S.}~\bibnamefont{Popescu}},
  \bibinfo{journal}{Phys. Rev. Lett.} \textbf{\bibinfo{volume}{80}},
  \bibinfo{pages}{1121} (\bibinfo{year}{1998}).

\bibitem[{\citenamefont{Furusawa et~al.}(1998)\citenamefont{Furusawa, Sørensen,
  Braunstein, Fuchs, Kimble, and Polzik}}]{Furusawa98}
\bibinfo{author}{\bibfnamefont{A.}~\bibnamefont{Furusawa}},
  \bibinfo{author}{\bibfnamefont{J.~L.} \bibnamefont{Sørensen}},
  \bibinfo{author}{\bibfnamefont{S.~L.} \bibnamefont{Braunstein}},
  \bibinfo{author}{\bibfnamefont{C.~A.} \bibnamefont{Fuchs}},
  \bibinfo{author}{\bibfnamefont{H.~J.} \bibnamefont{Kimble}},
  \bibnamefont{and} \bibinfo{author}{\bibfnamefont{E.~S.}
  \bibnamefont{Polzik}}, \bibinfo{journal}{Science}
  \textbf{\bibinfo{volume}{282}}, \bibinfo{pages}{706} (\bibinfo{year}{1998}).

\bibitem[{\citenamefont{Bowen et~al.}(2002)\citenamefont{Bowen, Treps,
  Schnabel, and Lam}}]{Bowen02}
\bibinfo{author}{\bibfnamefont{W.~P.} \bibnamefont{Bowen}},
  \bibinfo{author}{\bibfnamefont{N.}~\bibnamefont{Treps}},
  \bibinfo{author}{\bibfnamefont{R.}~\bibnamefont{Schnabel}}, \bibnamefont{and}
  \bibinfo{author}{\bibfnamefont{P.~K.} \bibnamefont{Lam}},
  \bibinfo{journal}{Phys. Rev. Lett.} \textbf{\bibinfo{volume}{89}},
  \bibinfo{pages}{253601} (\bibinfo{year}{2002}).

\bibitem[{\citenamefont{Zhang et~al.}(2003)\citenamefont{Zhang, Goh, Chou,
  Lodahl, and Kimble}}]{Zhang03}
\bibinfo{author}{\bibfnamefont{T.~C.} \bibnamefont{Zhang}},
  \bibinfo{author}{\bibfnamefont{K.~W.} \bibnamefont{Goh}},
  \bibinfo{author}{\bibfnamefont{C.~W.} \bibnamefont{Chou}},
  \bibinfo{author}{\bibfnamefont{P.}~\bibnamefont{Lodahl}}, \bibnamefont{and}
  \bibinfo{author}{\bibfnamefont{H.~J.} \bibnamefont{Kimble}},
  \bibinfo{journal}{Phys. Rev. A} \textbf{\bibinfo{volume}{67}},
  \bibinfo{pages}{033802} (\bibinfo{year}{2003}).

\bibitem[{\citenamefont{Knill et~al.}(2001)\citenamefont{Knill, Laflamme, and
  Milburn}}]{Knill01}
\bibinfo{author}{\bibfnamefont{E.}~\bibnamefont{Knill}},
  \bibinfo{author}{\bibfnamefont{R.}~\bibnamefont{Laflamme}}, \bibnamefont{and}
  \bibinfo{author}{\bibfnamefont{G.~J.} \bibnamefont{Milburn}},
  \bibinfo{journal}{Nature} \textbf{\bibinfo{volume}{409}}, \bibinfo{pages}{46}
  (\bibinfo{year}{2001}).

\bibitem[{\citenamefont{O'Brien et~al.}(2003)\citenamefont{O'Brien, Pryde,
  White, Ralph, and Branning}}]{OBrien03}
\bibinfo{author}{\bibfnamefont{J.~L.} \bibnamefont{O'Brien}},
  \bibinfo{author}{\bibfnamefont{G.~J.} \bibnamefont{Pryde}},
  \bibinfo{author}{\bibfnamefont{A.~G.} \bibnamefont{White}},
  \bibinfo{author}{\bibfnamefont{T.~C.} \bibnamefont{Ralph}}, \bibnamefont{and}
  \bibinfo{author}{\bibfnamefont{D.}~\bibnamefont{Branning}},
  \bibinfo{journal}{Nature} \textbf{\bibinfo{volume}{426}}, \bibinfo{pages}{264
  } (\bibinfo{year}{2003}).

\bibitem[{\citenamefont{Sanaka et~al.}(2004)\citenamefont{Sanaka, Jennewein,
  Pan, Resch, and Zeilinger}}]{Sanaka04}
\bibinfo{author}{\bibfnamefont{K.}~\bibnamefont{Sanaka}},
  \bibinfo{author}{\bibfnamefont{T.}~\bibnamefont{Jennewein}},
  \bibinfo{author}{\bibfnamefont{J.-W.} \bibnamefont{Pan}},
  \bibinfo{author}{\bibfnamefont{K.}~\bibnamefont{Resch}}, \bibnamefont{and}
  \bibinfo{author}{\bibfnamefont{A.}~\bibnamefont{Zeilinger}},
  \bibinfo{journal}{Phys. Rev. Lett.} \textbf{\bibinfo{volume}{92}},
  \bibinfo{pages}{017902} (\bibinfo{year}{2004}).

\bibitem[{\citenamefont{Boller et~al.}(1991)\citenamefont{Boller, Imamo\u{g}lu,
  and Harris}}]{Boller91}
\bibinfo{author}{\bibfnamefont{K.~J.} \bibnamefont{Boller}},
  \bibinfo{author}{\bibfnamefont{A.}~\bibnamefont{Imamo\u{g}lu}},
  \bibnamefont{and} \bibinfo{author}{\bibfnamefont{S.~E.}
  \bibnamefont{Harris}}, \bibinfo{journal}{Phys. Rev. Lett.}
  \textbf{\bibinfo{volume}{66}}, \bibinfo{pages}{2593} (\bibinfo{year}{1991}).

\bibitem[{\citenamefont{Arimondo}(1996)}]{Arimondo96}
\bibinfo{author}{\bibfnamefont{E.}~\bibnamefont{Arimondo}}, in
  \emph{\bibinfo{booktitle}{Progress in Optics}}, edited by
  \bibinfo{editor}{\bibfnamefont{E.}~\bibnamefont{Wolf}}
  (\bibinfo{publisher}{Elsevier Science}, \bibinfo{address}{Amsterdam},
  \bibinfo{year}{1996}), vol. \bibinfo{volume}{XXXV}, p. \bibinfo{pages}{257}.

\bibitem[{\citenamefont{Harris}(1997)}]{Harris97}
\bibinfo{author}{\bibfnamefont{S.~E.} \bibnamefont{Harris}},
  \bibinfo{journal}{Physics Today} \textbf{\bibinfo{volume}{50}},
  \bibinfo{pages}{36} (\bibinfo{year}{1997}).

\bibitem[{\citenamefont{Hau et~al.}(1999)\citenamefont{Hau, Harris, Dutton, and
  Berhoozi}}]{Hau99}
\bibinfo{author}{\bibfnamefont{L.~V.} \bibnamefont{Hau}},
  \bibinfo{author}{\bibfnamefont{S.~E.} \bibnamefont{Harris}},
  \bibinfo{author}{\bibfnamefont{Z.}~\bibnamefont{Dutton}}, \bibnamefont{and}
  \bibinfo{author}{\bibfnamefont{C.~H.} \bibnamefont{Berhoozi}},
  \bibinfo{journal}{Nature} \textbf{\bibinfo{volume}{397}},
  \bibinfo{pages}{594} (\bibinfo{year}{1999}).

\bibitem[{\citenamefont{Schmidt and Imamo\u{g}lu}(1996)}]{Schmidt96}
\bibinfo{author}{\bibfnamefont{H.}~\bibnamefont{Schmidt}} \bibnamefont{and}
  \bibinfo{author}{\bibfnamefont{A.}~\bibnamefont{Imamo\u{g}lu}},
  \bibinfo{journal}{Opt. Lett.} \textbf{\bibinfo{volume}{21}},
  \bibinfo{pages}{1936} (\bibinfo{year}{1996}).

\bibitem[{\citenamefont{Kang and Zhu}(2003)}]{Kang03}
\bibinfo{author}{\bibfnamefont{H.}~\bibnamefont{Kang}} \bibnamefont{and}
  \bibinfo{author}{\bibfnamefont{Y.}~\bibnamefont{Zhu}},
  \bibinfo{journal}{Phys. Rev. Lett.} \textbf{\bibinfo{volume}{91}},
  \bibinfo{pages}{093601} (\bibinfo{year}{2003}).

\bibitem[{\citenamefont{Imamo\u{g}lu et~al.}(1997)\citenamefont{Imamo\u{g}lu,
  Schmidt, Woods, and Deutsch}}]{Imamoglu97}
\bibinfo{author}{\bibfnamefont{A.}~\bibnamefont{Imamo\u{g}lu}},
  \bibinfo{author}{\bibfnamefont{H.}~\bibnamefont{Schmidt}},
  \bibinfo{author}{\bibfnamefont{G.}~\bibnamefont{Woods}}, \bibnamefont{and}
  \bibinfo{author}{\bibfnamefont{M.}~\bibnamefont{Deutsch}},
  \bibinfo{journal}{Phys. Rev. Lett.} \textbf{\bibinfo{volume}{79}},
  \bibinfo{pages}{1467} (\bibinfo{year}{1997}).

\bibitem[{\citenamefont{Grangier et~al.}(1998)\citenamefont{Grangier, Walls,
  and Gheri}}]{Grangier98}
\bibinfo{author}{\bibfnamefont{P.}~\bibnamefont{Grangier}},
  \bibinfo{author}{\bibfnamefont{D.~F.} \bibnamefont{Walls}}, \bibnamefont{and}
  \bibinfo{author}{\bibfnamefont{K.~M.} \bibnamefont{Gheri}},
  \bibinfo{journal}{Phys. Rev. Lett.} \textbf{\bibinfo{volume}{81}},
  \bibinfo{pages}{2833} (\bibinfo{year}{1998}).

\bibitem[{\citenamefont{Rebi\'{c} et~al.}(1999)\citenamefont{Rebi\'{c}, Tan,
  Parkins, and Walls}}]{Rebic99}
\bibinfo{author}{\bibfnamefont{S.}~\bibnamefont{Rebi\'{c}}},
  \bibinfo{author}{\bibfnamefont{S.~M.} \bibnamefont{Tan}},
  \bibinfo{author}{\bibfnamefont{A.~S.} \bibnamefont{Parkins}},
  \bibnamefont{and} \bibinfo{author}{\bibfnamefont{D.~F.} \bibnamefont{Walls}},
  \bibinfo{journal}{J. Opt. B: Quant. Semiclass. Opt.}
  \textbf{\bibinfo{volume}{1}}, \bibinfo{pages}{490} (\bibinfo{year}{1999}).

\bibitem[{\citenamefont{Gheri et~al.}(1999)\citenamefont{Gheri, Alge, and
  Grangier}}]{Gheri99}
\bibinfo{author}{\bibfnamefont{K.~M.} \bibnamefont{Gheri}},
  \bibinfo{author}{\bibfnamefont{W.}~\bibnamefont{Alge}}, \bibnamefont{and}
  \bibinfo{author}{\bibfnamefont{P.}~\bibnamefont{Grangier}},
  \bibinfo{journal}{Phys. Rev. A} \textbf{\bibinfo{volume}{60}},
  \bibinfo{pages}{R2673} (\bibinfo{year}{1999}).

\bibitem[{\citenamefont{Greentree et~al.}(2000)\citenamefont{Greentree,
  Vaccaro, de~Echaniz, Durant, and Marangos}}]{Greentree00}
\bibinfo{author}{\bibfnamefont{A.~D.} \bibnamefont{Greentree}},
  \bibinfo{author}{\bibfnamefont{J.~A.} \bibnamefont{Vaccaro}},
  \bibinfo{author}{\bibfnamefont{S.~R.} \bibnamefont{de~Echaniz}},
  \bibinfo{author}{\bibfnamefont{A.~V.} \bibnamefont{Durant}},
  \bibnamefont{and} \bibinfo{author}{\bibfnamefont{J.~P.}
  \bibnamefont{Marangos}}, \bibinfo{journal}{J. Opt. B: Quantum Semiclass.
  Opt.} \textbf{\bibinfo{volume}{2}}, \bibinfo{pages}{252}
  (\bibinfo{year}{2000}).

\bibitem[{\citenamefont{Roch et~al.}(1997)\citenamefont{Roch, Vigneron, Grelu,
  Sinatra, Poizat, and Grangier}}]{Roch97}
\bibinfo{author}{\bibfnamefont{J.-F.} \bibnamefont{Roch}},
  \bibinfo{author}{\bibfnamefont{K.}~\bibnamefont{Vigneron}},
  \bibinfo{author}{\bibfnamefont{P.}~\bibnamefont{Grelu}},
  \bibinfo{author}{\bibfnamefont{A.}~\bibnamefont{Sinatra}},
  \bibinfo{author}{\bibfnamefont{J.-P.} \bibnamefont{Poizat}},
  \bibnamefont{and} \bibinfo{author}{\bibfnamefont{P.}~\bibnamefont{Grangier}},
  \bibinfo{journal}{Phys. Rev. Lett.} \textbf{\bibinfo{volume}{78}},
  \bibinfo{pages}{634} (\bibinfo{year}{1997}).

\bibitem[{\citenamefont{Wang et~al.}(2001)\citenamefont{Wang, Goorskey, and
  Xiao}}]{Wang01}
\bibinfo{author}{\bibfnamefont{H.}~\bibnamefont{Wang}},
  \bibinfo{author}{\bibfnamefont{D.}~\bibnamefont{Goorskey}}, \bibnamefont{and}
  \bibinfo{author}{\bibfnamefont{M.}~\bibnamefont{Xiao}},
  \bibinfo{journal}{Phys. Rev. Lett.} \textbf{\bibinfo{volume}{87}},
  \bibinfo{pages}{073601} (\bibinfo{year}{2001}).

\bibitem[{\citenamefont{Zubairy et~al.}(2002)\citenamefont{Zubairy, Matsko, and
  Scully}}]{Zubairy02}
\bibinfo{author}{\bibfnamefont{M.~S.} \bibnamefont{Zubairy}},
  \bibinfo{author}{\bibfnamefont{A.~B.} \bibnamefont{Matsko}},
  \bibnamefont{and} \bibinfo{author}{\bibfnamefont{M.~O.}
  \bibnamefont{Scully}}, \bibinfo{journal}{Phys. Rev. A}
  \textbf{\bibinfo{volume}{65}}, \bibinfo{pages}{043804}
  (\bibinfo{year}{2002}).

\bibitem[{\citenamefont{Matsko et~al.}(2002)\citenamefont{Matsko, Novikova,
  Welch, and Zubairy}}]{Matsko02}
\bibinfo{author}{\bibfnamefont{A.~B.} \bibnamefont{Matsko}},
  \bibinfo{author}{\bibfnamefont{I.}~\bibnamefont{Novikova}},
  \bibinfo{author}{\bibfnamefont{G.~R.} \bibnamefont{Welch}}, \bibnamefont{and}
  \bibinfo{author}{\bibfnamefont{M.~S.} \bibnamefont{Zubairy}},
  \bibinfo{journal}{Opt. Lett.} \textbf{\bibinfo{volume}{28}},
  \bibinfo{pages}{96} (\bibinfo{year}{2002}).

\bibitem[{\citenamefont{Matsko et~al.}(2003)\citenamefont{Matsko, Novikova,
  Zubairy, and Welch}}]{Matsko03}
\bibinfo{author}{\bibfnamefont{A.~B.} \bibnamefont{Matsko}},
  \bibinfo{author}{\bibfnamefont{I.}~\bibnamefont{Novikova}},
  \bibinfo{author}{\bibfnamefont{M.~S.} \bibnamefont{Zubairy}},
  \bibnamefont{and} \bibinfo{author}{\bibfnamefont{G.~R.} \bibnamefont{Welch}},
  \bibinfo{journal}{Phys. Rev. A} \textbf{\bibinfo{volume}{67}},
  \bibinfo{pages}{043805} (\bibinfo{year}{2003}).

\bibitem[{\citenamefont{Greentree et~al.}(2003)\citenamefont{Greentree,
  Richards, Vaccaro, Durrant, de~Echaniz, Segal, and Marangos}}]{Greentree03}
\bibinfo{author}{\bibfnamefont{A.~D.} \bibnamefont{Greentree}},
  \bibinfo{author}{\bibfnamefont{D.}~\bibnamefont{Richards}},
  \bibinfo{author}{\bibfnamefont{J.~A.} \bibnamefont{Vaccaro}},
  \bibinfo{author}{\bibfnamefont{A.~V.} \bibnamefont{Durrant}},
  \bibinfo{author}{\bibfnamefont{S.~R.} \bibnamefont{de~Echaniz}},
  \bibinfo{author}{\bibfnamefont{D.~M.} \bibnamefont{Segal}}, \bibnamefont{and}
  \bibinfo{author}{\bibfnamefont{J.~P.} \bibnamefont{Marangos}},
  \bibinfo{journal}{Phys. Rev. A} \textbf{\bibinfo{volume}{67}},
  \bibinfo{pages}{023818} (\bibinfo{year}{2003}).

\bibitem[{\citenamefont{Ottaviani et~al.}(2003)\citenamefont{Ottaviani, Vitali,
  Artoni, Cataliotti, and Tombesi}}]{Ottaviani03}
\bibinfo{author}{\bibfnamefont{C.}~\bibnamefont{Ottaviani}},
  \bibinfo{author}{\bibfnamefont{D.}~\bibnamefont{Vitali}},
  \bibinfo{author}{\bibfnamefont{M.}~\bibnamefont{Artoni}},
  \bibinfo{author}{\bibfnamefont{F.}~\bibnamefont{Cataliotti}},
  \bibnamefont{and} \bibinfo{author}{\bibfnamefont{P.}~\bibnamefont{Tombesi}},
  \bibinfo{journal}{Phys. Rev. Lett.} \textbf{\bibinfo{volume}{90}},
  \bibinfo{pages}{197902} (\bibinfo{year}{2003}).

\bibitem[{\citenamefont{Lloyd}(1995)}]{Lloyd95}
\bibinfo{author}{\bibfnamefont{S.}~\bibnamefont{Lloyd}},
  \bibinfo{journal}{Phys. Rev. Lett.} \textbf{\bibinfo{volume}{75}},
  \bibinfo{pages}{346} (\bibinfo{year}{1995}).

\bibitem[{\citenamefont{Lukin and Imamo\u{g}lu}(2000)}]{Lukin00}
\bibinfo{author}{\bibfnamefont{M.~D.} \bibnamefont{Lukin}} \bibnamefont{and}
  \bibinfo{author}{\bibfnamefont{A.}~\bibnamefont{Imamo\u{g}lu}},
  \bibinfo{journal}{Phys. Rev. Lett.} \textbf{\bibinfo{volume}{84}},
  \bibinfo{pages}{1419} (\bibinfo{year}{2000}).

\bibitem[{\citenamefont{Lukin and Imamo\u{g}lu}(2001)}]{Lukin01}
\bibinfo{author}{\bibfnamefont{M.~D.} \bibnamefont{Lukin}} \bibnamefont{and}
  \bibinfo{author}{\bibfnamefont{A.}~\bibnamefont{Imamo\u{g}lu}},
  \bibinfo{journal}{Nature} \textbf{\bibinfo{volume}{413}},
  \bibinfo{pages}{273} (\bibinfo{year}{2001}).

\bibitem[{\citenamefont{Petrosyan and Kurizki}(2002)}]{Petrosyan02}
\bibinfo{author}{\bibfnamefont{D.}~\bibnamefont{Petrosyan}} \bibnamefont{and}
  \bibinfo{author}{\bibfnamefont{G.}~\bibnamefont{Kurizki}},
  \bibinfo{journal}{Phys. Rev. A} \textbf{\bibinfo{volume}{65}},
  \bibinfo{pages}{033833} (\bibinfo{year}{2002}).

\bibitem[{\citenamefont{Turchette et~al.}(1995)\citenamefont{Turchette, Hood,
  Lange, Mabuchi, and Kimble}}]{Turchette95}
\bibinfo{author}{\bibfnamefont{Q.~A.} \bibnamefont{Turchette}},
  \bibinfo{author}{\bibfnamefont{C.~J.} \bibnamefont{Hood}},
  \bibinfo{author}{\bibfnamefont{W.}~\bibnamefont{Lange}},
  \bibinfo{author}{\bibfnamefont{H.}~\bibnamefont{Mabuchi}}, \bibnamefont{and}
  \bibinfo{author}{\bibfnamefont{H.~J.} \bibnamefont{Kimble}},
  \bibinfo{journal}{Phys. Rev. Lett.} \textbf{\bibinfo{volume}{75}},
  \bibinfo{pages}{4710} (\bibinfo{year}{1995}).

\bibitem[{\citenamefont{Resch et~al.}(2002)\citenamefont{Resch, Lundeen, and
  Steinberg}}]{Resch02}
\bibinfo{author}{\bibfnamefont{K.~J.} \bibnamefont{Resch}},
  \bibinfo{author}{\bibfnamefont{J.~S.} \bibnamefont{Lundeen}},
  \bibnamefont{and} \bibinfo{author}{\bibfnamefont{A.~M.}
  \bibnamefont{Steinberg}}, \bibinfo{journal}{Phys. Rev. Lett.}
  \textbf{\bibinfo{volume}{89}}, \bibinfo{pages}{037904}
  (\bibinfo{year}{2002}).

\bibitem[{\citenamefont{Unanyan et~al.}(1998)\citenamefont{Unanyan,
  Fleischhauer, Shore, and Bergmann}}]{Unanyan98}
\bibinfo{author}{\bibfnamefont{R.}~\bibnamefont{Unanyan}},
  \bibinfo{author}{\bibfnamefont{M.}~\bibnamefont{Fleischhauer}},
  \bibinfo{author}{\bibfnamefont{B.~W.} \bibnamefont{Shore}}, \bibnamefont{and}
  \bibinfo{author}{\bibfnamefont{K.}~\bibnamefont{Bergmann}},
  \bibinfo{journal}{Opt. Comm.} \textbf{\bibinfo{volume}{155}},
  \bibinfo{pages}{144} (\bibinfo{year}{1998}).

\bibitem[{\citenamefont{Paspalakis and
  Knight}(2002{\natexlab{a}})}]{Paspalakis02a}
\bibinfo{author}{\bibfnamefont{E.}~\bibnamefont{Paspalakis}} \bibnamefont{and}
  \bibinfo{author}{\bibfnamefont{P.~L.} \bibnamefont{Knight}},
  \bibinfo{journal}{J. Opt. B: Quantum Semiclass. Opt.}
  \textbf{\bibinfo{volume}{4}}, \bibinfo{pages}{S372}
  (\bibinfo{year}{2002}{\natexlab{a}}).

\bibitem[{\citenamefont{Paspalakis et~al.}(2002)\citenamefont{Paspalakis,
  Kylstra, and Knight}}]{Paspalakis02b}
\bibinfo{author}{\bibfnamefont{E.}~\bibnamefont{Paspalakis}},
  \bibinfo{author}{\bibfnamefont{N.~J.} \bibnamefont{Kylstra}},
  \bibnamefont{and} \bibinfo{author}{\bibfnamefont{P.~L.}
  \bibnamefont{Knight}}, \bibinfo{journal}{Phys. Rev. A}
  \textbf{\bibinfo{volume}{65}}, \bibinfo{pages}{053808}
  (\bibinfo{year}{2002}).

\bibitem[{\citenamefont{Paspalakis and
  Knight}(2002{\natexlab{b}})}]{Paspalakis02c}
\bibinfo{author}{\bibfnamefont{E.}~\bibnamefont{Paspalakis}} \bibnamefont{and}
  \bibinfo{author}{\bibfnamefont{P.~L.} \bibnamefont{Knight}},
  \bibinfo{journal}{J. Mod. Opt.} \textbf{\bibinfo{volume}{49}},
  \bibinfo{pages}{87} (\bibinfo{year}{2002}{\natexlab{b}}).

\bibitem[{\citenamefont{Malakyan}()}]{Malakyan01}
\bibinfo{author}{\bibfnamefont{Y.~P.} \bibnamefont{Malakyan}},
  \bibinfo{note}{e-print, quant-ph/0112058}.

\bibitem[{\citenamefont{Petrosyan and Malakyan}()}]{Petrosyan04}
\bibinfo{author}{\bibfnamefont{D.}~\bibnamefont{Petrosyan}} \bibnamefont{and}
  \bibinfo{author}{\bibfnamefont{Y.~P.} \bibnamefont{Malakyan}},
  \bibinfo{note}{e-print, quant-ph/0402070}.

\bibitem[{\citenamefont{Walls and Milburn}(1994)}]{Walls94}
\bibinfo{author}{\bibfnamefont{D.~F.} \bibnamefont{Walls}} \bibnamefont{and}
  \bibinfo{author}{\bibfnamefont{G.~J.} \bibnamefont{Milburn}},
  \emph{\bibinfo{title}{Quantum Optics}} (\bibinfo{publisher}{Springer},
  \bibinfo{address}{Berlin}, \bibinfo{year}{1994}).

\end{thebibliography}
\end{document}